\documentclass[a4paper,12pt]{article}
\usepackage[round]{natbib}
\usepackage[english]{babel}
\usepackage{setspace}
\usepackage[utf8x]{inputenc}
\usepackage{graphicx} 
\usepackage[left=2cm,right=2cm,top=2cm,bottom=2cm]{geometry}
\usepackage{textcomp}
\usepackage[T1]{fontenc}
\usepackage{amssymb}
\usepackage{xcolor}
\usepackage{authblk}
\usepackage{ulem}
\usepackage{hyperref}

\title{
\large
\textbf{Mesoscale modeling of Venus' bow-shape waves}
}

\author[1]{Maxence Lef{\`e}vre}
\author[1,2]{Aymeric Spiga}
\author[1]{S{\'e}bastien Lebonnois}
\affil[1]{Laboratoire de M\'et\'eorologie Dynamique (LMD/IPSL), Sorbonne Universit{\'e}, Centre National de la Recherche Scientifique, {\'E}cole Polytechnique, {\'E}cole Normale Sup{\'e}rieure, Paris, France}
\affil[2]{Institut Universitaire de France, Paris, France}

\date{Accpeted in Icarus \href{https://doi.org/10.1016/j.icarus.2019.07.010}{DOI}}

\begin{document}

\maketitle
\newpage
\section*{Abstract}

The \textit{Akatsuki} instrument LIR measured an unprecedented wave feature at the top of Venusian cloud layer. Stationary bow-shape waves of thousands of kilometers large lasting several Earth days have been observed over the main equatorial mountains. Here we use for the first time a mesoscale model of the Venus's atmosphere with high-resolution topography and fully coupled interactive radiative transfer computations. Mountain waves resolved by the model form large-scale bow shape waves with an amplitude of about 1.5~K and a size up to several decades of latitude similar to the ones measured by the \textit{Akatsuki} spacecraft. The maximum amplitude of the waves appears in the afternoon due to an increase of the near-surface stability. Propagating vertically the waves encounter two regions of low static stability, the mixed layer between approximately 18 and 30~km and the convective layer between 50 and 55~km. Some part of the wave energy can pass through these regions via wave tunneling. These two layers act as wave filter, especially the deep atmosphere layer. The encounter with these layers generates trapped lee waves propagating horizontally. No stationary waves is resolved at cloud top over the polar regions because of strong circumpolar transient waves, and a thicker deep atmosphere mixed layer that filters most of the mountain waves. 

\section{Intro}

The influence of the topography on the Venusian atmosphere, especially on the dynamics of the cloud layer, is still not fully understood. The \textit{VeGa} balloons campaign was the first measurement of the impact of the surface on the atmospheric dynamics and demonstrated an increase of the vertical wind above Aphrodite Terra, speculated to be linked to the propagation of orographic gravity waves \citep{Blam86}. Numerical modeling was then used to assess the possibility for such kind of wave to emerge \cite{Youn87,Youn94}. These modeling efforts showed that gravity waves generated by the topography can propagate up to the cloud layer, even in the observed conditions of strong vertical variations of zonal wind and stability. Additional observations of the interactions between the surface and the atmosphere have been made with the \textit{Venus Express} mission \citep{Bert16}. Correlations between the zonal wind at the top of the cloud and the underlying topography has been evidenced by UV measurements with the Venus Monitoring Camera, and interpreted as the result of stationary gravity waves. A water minimum has also been measured with \textit{Venus Express} instrument SPICAV at cloud top above Aphrodite Terra \citep{Fedo16} possibly linked to an interaction between the surface and the cloud layer.

Using the LIR instrument on board the \textit{Akatsuki} mission, \cite{Fuku17} discovered large-scale stationary bow-shaped oscillations at the top of the cloud above Aphrodite Terra. This bow-shape signature, extending over 60\textdegree~of latitude (about 10 000~km across), was observed during 5 days. Similar signatures were also reported above the main equatorial topographic features \citep[e.g., Atla and Beta Regio]{Kouy17}. Those signatures were interpreted as stationary orographic gravity waves. The bow-shape wave above Aphrodite Terra has been observed in the afternoon, with a maximum amplitude close to the evening terminator. The cloud-top signatures above the other Venusian mountains are visible during the afternoon, with a maximum close to the evening terminator. Cloud-top signatures associated with the underlying topography were also observed above Beta Regio with the instrument \textit{Akatsuki}/IR2 \citep[2.02~$\mu$m wavelength,][]{Sato17} as well as with \textit{Akatsuki} UV imager above Aphrodite Terra, Atla and Beta Regio \citep{Kita19}. Stationary features have also been measured by the \textit{Venus Express} instrument VIRTIS above the main topographical obstacles in the southern hemisphere in the nightside with no apparent dependence with the local time \citep{Pera17}. 

To investigate the atmospheric dynamics of those observed phenomena, two modeling approaches were adopted: one by \cite{Fuku17} using a very idealized (i.e. no representation of physical processes and topography) high-resolution Global Circulation Model (GCM), with an arbitrary perturbation as the source of the waves and one by \cite{Nava18} using a subgrid-scale paramaterization for gravity waves in a low-resolution GCM with complete physics \citep[the IPSL Venus GCM described in][]{Gara18}. In the study by \cite{Nava18}, the parametric representation of the orographic gravity waves in a GCM demonstrated the link between the topography and the bow-shaped waves. In addition, the impact of the large-scale mountain waves detected by Akatsuki on the global-scale dynamics was shown to induce, along with the thermal tide and baroclinic waves, a significant change in the rotation rate of the solid body, with variability of the order of minutes.

Despite those recent modeling studies, open questions remain on the source, propagation and impact of the large-scale gravity waves evidenced by Akatsuki \citep{Fuku17,Kouy17}. Notably, \cite{Nava18} used a parameterization for orographic gravity waves on Venus and did not resolve the complete emission and propagation of gravity waves from the surface to Venus' cloud top, which is not possible with current GCMs for Venus. Following the methodology described in \cite{Lefe18} to study the turbulence and gravity waves in the cloud layers, we propose a method combining high-resolution atmospheric dynamics with detailed physics of the Venusian atmosphere to resolve the emission and propagation of mountain waves in a realistic atmosphere. In other words, in order to address the Venusian bow-shaped mountain waves, we developed the first Venusian mesoscale model. With high-resolution topography from \textit{Magellan} data, we can model the atmospheric dynamics of a limited area of Venus using the dynamical core of the Weather Research and Forecast \citep[WRF,][]{Skam08}. Following a method used on Mars by \cite{Spig09}, the WRF dynamical core is interfaced with the physics of the IPSL Venus GCM \citep{Lebo10,Gara18} to be able to compute radiative transfer and subgrid-scale turbulent mixing coupled in real-time with the dynamical integrations. 

In this paper, using our new mesoscale model for Venus, we carry out simulations in areas surrounding the main mountains of the Venusian equatorial belt at various local times, in order to understand the generation and vertical propagation of the waves and to interpret the signal detected by Akatsuki at the top of the cloud. Our new mesoscale model for Venus can potentially be used for many applications other than studying the bow-shaped mountain waves, e.g. near-surface slope winds \citep{Lebo18}, polar meteorology \cite{Gara15}, mesoscale structures in the vicinity of the super-rotating jet \citep{Hori17}. 

This paper is organized as follows. Our mesoscale model for Venus is described in Section 2. In Section 3, the emission of resolved gravity waves is presented. The resulting top-of-the-cloud signal above Aphrodite Terra is developed in Section 4, and above Atla Regio and Beta Region in Section 5. In Section 6, mountain waves in the polar regions are discussed. Our conclusions are summarized in Section 7. 
 
\section{The LMD Venus mesoscale model}
\label{Sec:model}

\subsection{Dynamical core}

The dynamical core of our LMD Venus mesoscale model is based on the Advanced Research Weather-Weather Research and Forecast (hereafter referred to as WRF) terrestrial model \citep{Skam08}. This methodology is similar to the one adopted for the LMD Mars mesoscale model (see \cite{Spig09} for a reference publication of this model and \cite{Spig18} for the most up-to-date description). The WRF dynamical core integrates the fully compressible non-hydrostatic Navier-Stokes equations on a defined area of the planet. The conservation of the mass, momentum, and entropy is ensured by an explicitly conservative flux-form formulation of the fundamental equations \citep{Skam08}, based on mass-coupled meteorological variables (winds and potential temperature). To ensure the stability of the model, and given the typical horizontal scales aimed at in our studies, the fundamental equations are integrated under the hydrostatic approximation, available as a runtime option in WRF. 

\subsection{Coupling with complete physical packages for Venus}

The radiative heating rates, solar and IR, are calculated using the IPSL Venus GCM radiative transfer scheme \citep{Lebo15}. This setting is similar to the ``online'' mode described in \cite{Lefe18}. The time step ratio between the dynamical and physical integrations is set to 250, as a trade-off between computational efficiency and the requirement that the physical timestep is significantly less than the typical radiative timescale for Venus.

The version of the Venus radiative transfer model used in our mesoscale model is the same as in the version of the IPSL Venus GCM described in \cite{Gara18}. The infrared (IR) transfer uses \cite{Eyme09} net-exchange rate (NER) formalism: the exchanges of energy between the layers are computed before the dynamical simulations are carried out by separating temperature-independent coefficients from the temperature-dependent Planck functions of the different layers. These temperature-independent coefficients are then used in the mesoscale simulations to compute the infrared cooling rates of each layer. The solar rates are based on \cite{Haus15} computations: look-up tables of vertical profiles of the solar heating rate as a function of solar zenith angle are read, before being interpolated on the vertical grid adopted for mesoscale integrations. 

The cloud model used in this study is based on the \cite{Haus14} and \cite{Haus15} models derived from retrievals carried out with \textit{Venus Express} instruments. The cloud structure is latitude-dependent, the cloud top varies from 71~km to 62~km between the Equator and the poles \cite{Haus14}. This latitudinal variation of the cloud takes the form of five distinct latitude intervals: 0$^{\circ}$ to 50$^{\circ}$, 50$^{\circ}$ to 60$^{\circ}$, 60$^{\circ}$ to 70$^{\circ}$, 70$^{\circ}$ to 80$^{\circ}$ and 80$^{\circ}$ to 90$^{\circ}$. For each latitudinal intervals different NER-coefficient matrices are computed on the vertical levels of the model, ranging from the surface to roughly 100~km altitude. The cloud structure used for the calculations is fixed prior to the simulation and does not evolve with time. 

Since the horizontal grid spacing for the mesoscale simulations is set to several tens of kilometer, the convective turbulence in the planetary boundary layer and the cloud layer is not resolved. Therefore, similarly to what is done in GCMs, our mesoscale model uses subgrid-scale parameterizations for turbulent mixing. As in the IPSL Venus GCM, for mixing by smaller-scale turbulent eddies, the formalism of \cite{Mell82} is adopted, which calculates with a prognostic equation the turbulent kinetic energy and mixing length. For mixing by larger-scale turbulent plumes \citep[such as those resolved by Large-Eddy Simulations, see][]{Lefe17,Lefe18}, a simple dry convective adjustment is used to compute mixed layers in situations of convectively-unstable temperature profiles. In the physics (same as in the GCM), the dependency of the heat capacity with temperature $c_p(T)$ is taken into account, but in the dynamical core, we use a constant heat capacity as in \citet{Lefe18}, with a mean value of $c_p = 1000$~J~kg$^{-1}$~K$^{-1}$ suitable for the vertical extent of our model from the surface to 100~km altitude. The initial state and the boundary conditions of the mesoscale domain, detailed in the next section, are calculated using a variable $c_p$ to ensure a realistic forcing of the mesoscale model by the large-scale dynamics. Therefore the impact of the constant value of $c_p$ is minimum in our configuration because the equations of the WRF dynamical core are formulated in potential temperature and the conversion from the dynamics to the physics (and vice versa) between potential temperature and temperature is done with a variable $c_p (T)$.

\subsection{Simulation settings}

\begin{figure}[!ht]
 \center
 \includegraphics[width=10cm]{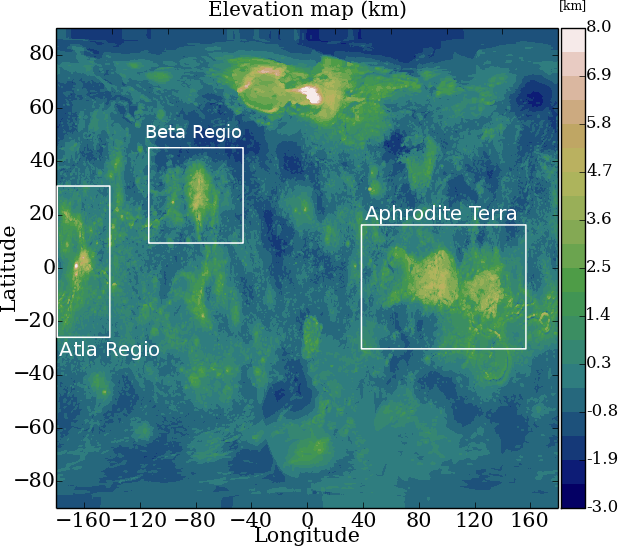}
 \caption{Topographic map of Venus used for the mesoscale simulations, from \textit{Magellan} \citep{Ford92} and \textit{Pioneer Venus} \citep{Pette80} datasets. The different domains used in the mesoscale simulations are indicated : Aphrodite Terra, Beta Regio and Atla Terra.}
 \label{fig:231}
\end{figure}

The topography used in our mesoscale model is presented in Figure~\ref{fig:231}, it is based on \textit{Magellan} data \citep{Ford92} for the majority of the surface, with \textit{Pioneer Venus} data used to fill the blank spots wherever needed \citep{Pette80}. The dataset has a resolution of 8192 points along the longitude and 4096 points along the latitude. Given the detection of bow-shaped features by the \textit{Akatsuki} spacecraft \citep{Kouy17}, we choose three domains of interest : Aphrodite Terra, Atla Regio and Beta Regio. These three mesoscale domains has been computed with a horizontal resolution of 40~km for Aphrodite Terra, 30~km for Beta Region and 15~km for Atla Regio. Given the horizontal resolutions involved, the dynamical timestep for mesoscale integrations is set between~8 and~10~s \citep[typical for those horizontal resolutions, see][]{Spig09}. The horizontal domains and resolutions have been chosen to enclose the whole latitudinal extent of the bow-shaped waves, while keeping a feasible computing time. The vertical grid is composed of 150 levels from the ground to 100~km, with a similar distribution as the LMD Venus LES model \citep{Lefe18}.  

The horizontal boundary conditions have been chosen as `specified', i.e. meteorological fields are extracted from IPSL Venus GCM simulations \citep[detailed in][]{Gara18} and interpolated both on the vertical grid, accounting for the refined topography of the mesoscale, and on the temporal dimension, accounting for the evolution of those fields over the low dynamical timestep in mesoscale integrations. This ensures that the planetary-scale super-rotation of the Venusian atmosphere is well represented. The mesoscale simulations presented in this study are performed using an update frequency of a 1/100 Venus day, which enables a correct representation of the large-scale variability simulated by the GCM at the boundaries of the mesoscale domain. Between the mesoscale domain and the specified boundary fields, a relaxation zone is implemented in order to allow for the development of the mesoscale circulations inside the domain, while keeping prescribed GCM fields at the boundaries \citep{Skam08}. In this study, the number of relaxation grid points is set to 5. An exponential function is used to reach a smooth transition between the domain and the specified boundary fields; the coefficient is set to 1, a value used for terrestrial and Martian applications that appeared suitable for the Venusian case as well. At the top of the mesoscale model, a diffusive Rayleigh damping layer is applied to avoid any spurious reflection of vertically-propagating gravity waves. The damping coefficient is set to 0.01 and the depth of the damping layer to 5~km.

The initialization of the meteorological fields for the mesoscale domain are, similarly to boundary conditions, extracted from the \cite{Gara18} GCM. First the fields are interpolated horizontally to the mesoscale refined-resolution grid, then interpolated vertically on the mesoscale vertical grid -- accounting for high-resolution topography. To extrapolate the high-resolution features, the same methodology is used as in the LMD Mars Mesoscale Model \citep{Spig09}. 

Venus's rotation is retrograde, but the WRF dynamical core has been built for Earth applications, and therefore assuming implicitly prograde rotation. In order to solve this issue, as is the case for the IPSL GCM runs \citep{Lebo10}, the GCM fields are turned upside down and the meridional wind is multiplied by minus 1. At the post-processing stage, the fields are turned upside down again and the meridional wind is again multiplied by minus 1, to obtain the diagnostics enclosed in this paper.

\section{Generation and propagation of orographic gravity waves}
\label{Sec:gwg}

\begin{figure}[!ht]
 \center
 \includegraphics[width=10cm]{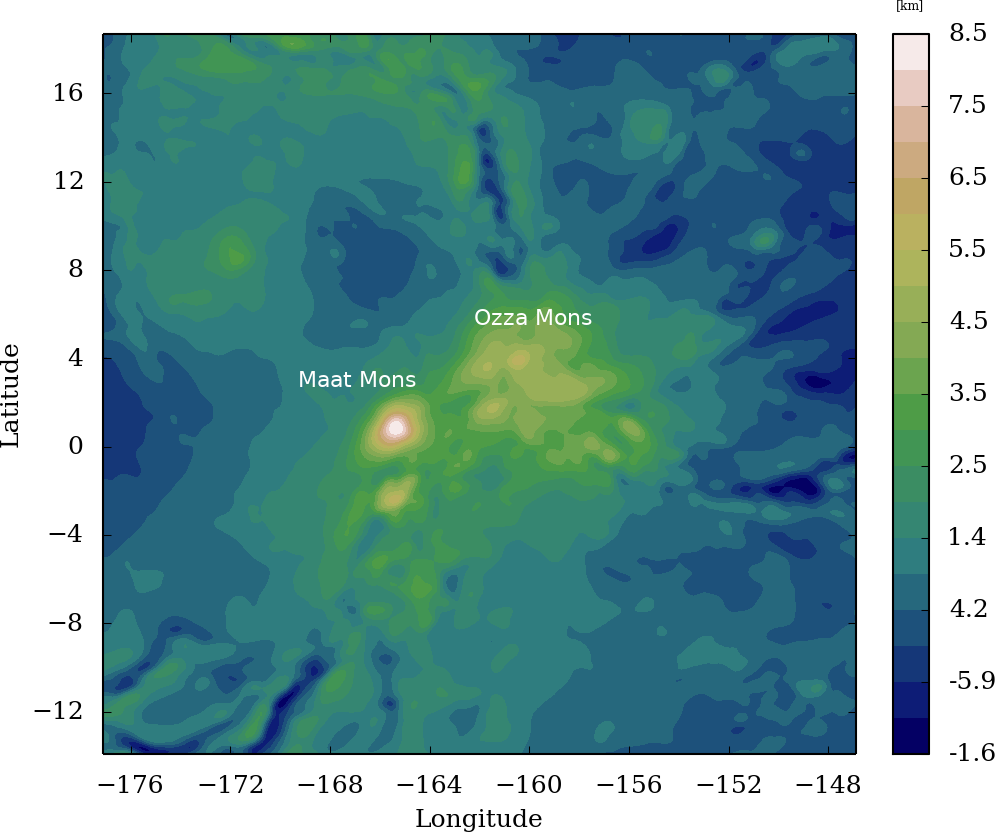}
 \caption{Elevation map (km) of the selected domain for Atla Regio with a resolution of 15~km.}
 \label{fig:31}
\end{figure}

To discuss the orographic wave generation, we focus first on Atla Regio which facilitates the visualization of the phenomena given its very sharp mountains. Figure~\ref{fig:31} shows the domain chosen for Atla Regio. 

\begin{figure}[!ht]
 \center
 \includegraphics[width=10cm]{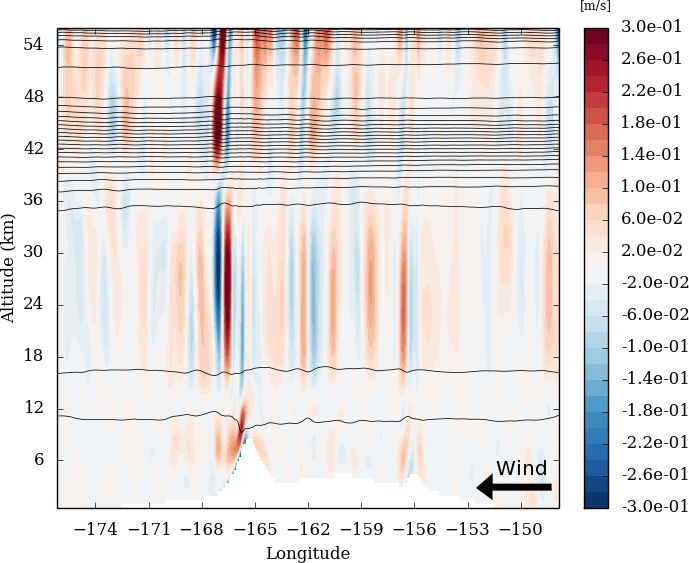}
 \caption{Vertical cross-section of the vertical wind (m~s$^{-1}$) at 1\textdegree~of latitude, between the surface and approximately 55~km in the beginning of afternoon. Contours represent the potential temperature. Direction of the zonal is indicated with the black arrow.}
 \label{fig:32}
\end{figure}

Figure~\ref{fig:32} shows a vertical cross-section of the vertical wind (m~s$^{-1}$) at 1\textdegree~of latitude, between the surface and approximately 55~km in the beginning of afternoon. Contours represent potential temperature. For this figure~\ref{fig:32}, and all the similar cross-sections shown in this paper, the zonal wind comes for the right side of the plot, i.e. wind is blowing westward. The Pioneer Venus probe measured horizontal wind (u and v) from the ground to 3~km between 0.1 and 1.5 m~s$^{-1}$ \citep{Schu80}. In the GCM, the horizontal wind on the same vertical extent is between 0.1 and about 3.0~m~s$^{-1}$. The horizontal wind is of the same order of magnitude as in-situ measurements although slightly overestimated. 

Strong vertical winds are visible above the two main mountains, Maat Mons at -165\textdegree~of longitude and Ozza Mons at -156\textdegree~of longitude. The interaction between the incoming zonal flow and those sharp topographical obstacles leads to the generation of gravity waves. Near the surface, the value of the dimensionless mountain height $H_d = H \, N_B / u$ can reach value close to 1, where $H$ is the maximum mountain height, $N_B$ is Brunt-V\"ais\"al\"a frequency and $u$ the zonal wind, meaning that the flow is in a non-linear regime \citep{Durr03}. The waves propagate vertically and encounter two regions of low static stability, the mixed layer between 18 and 35~km and the convective layer between 48 and 52~km. Above, the waves propagate into the stratified layers. The vertical wavelength of the waves is around 30~km.

\begin{figure}[!ht]
 \center
 \includegraphics[width=8cm]{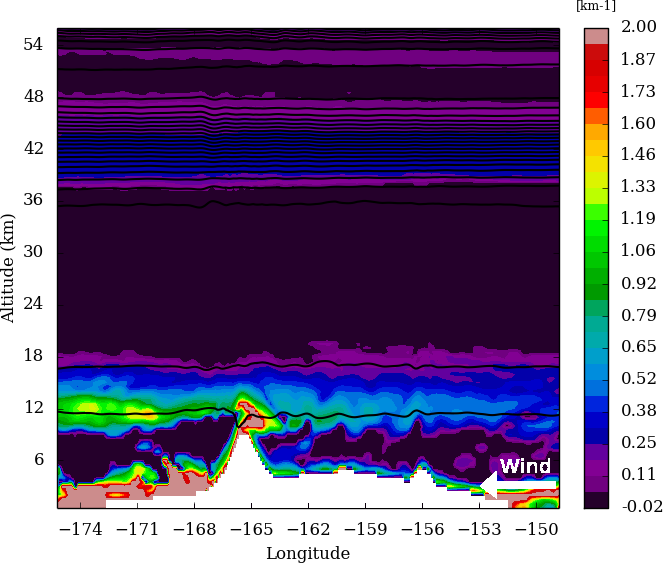}
 \includegraphics[width=8cm]{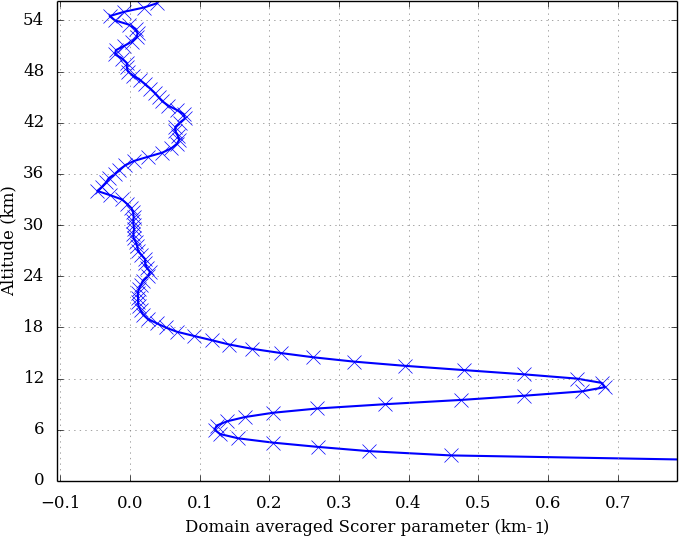}
 \caption{Left : Instantaneous vertical cross-section of the Scorer parameter (km$^{-1}$) at 1\textdegree~of latitude, between the surface and approximately 55~km in the beginning of afternoon. Contours represent the potential temperature. Right : domain-averaged vertical profile of the Scorer parameter at the same local time. Direction of the zonal is indicated with the white arrow.}
 \label{fig:33}
\end{figure}

The impact of these two near-neutral low-stability regions on the vertical propagation of the wave can be understood via the Scorer parameter (km$^{-1}$) \citep{Scor49} : $l^2 = \frac{{N_B}^2}{u^2} - \frac{1}{u}\frac{d^2 u}{d z^2}$. This parameter represents the minimum vertical wavelength of propagation. Figure~\ref{fig:33} shows the vertical cross-section of the Scorer parameter, and its domain-averaged vertical profile, at the same location as Figure.~\ref{fig:32}. 

Beyond 12~km altitude, the Scorer parameter decreases strongly up to 18~km and has very small values, sometimes negative, from 18 to 35~km. Then it increases up to 42~km, and decreases until reaching the convective layer at 48~km. Inside this convective layer, the Scorer parameter is again very small. Above the convective layer, the Scorer parameter increases through the stable atmosphere. These two regions of low Scorer parameter indicate the predominance of trapped lee waves that propagate horizontally. The horizontally-propagating trapped waves are visible in Figure~\ref{fig:32} within layers exhibiting low Scorer parameter, between 18 and 35~km and between 48 and around 54~km. An additional vertical propagation of those trapped waves is visible at some longitudes. This phenomenon is similar to the leakage of trapped lee waves into the stratosphere on Earth \citep{Brow83}. On Venus, the horizontal wavelength is approximately 150~km, one order of magnitude larger to typical Earth's trapped lee waves \citep{Ralp97}. Large horizontal wavelengths are known to increase the vertical leakage of trapped lee waves \citep{Durr15}. The horizontal wavelength of the trapped lee waves generated above Atla Regio is consistent with VEx/VIRTIS nightside measurements in the low cloud region \citep{Pera08}, whereas the trapped lee waves generated above Aphrodite Terra and Beta Regio are larger by at least a factor of 2.

\bigbreak
One of the main questions about the vertical propagation of bow-shaped waves on Venus (and the reason why they can be detected at high altitudes by Akatsuki) is how the waves can propagate through two (near-neutral) mixed layers. A similar configuration can be found in the Earth's oceans where the seasonal and main thermoclines can be separated by a relatively weakly stratified region \citep{Ecka61}. An analogous case in the Earth atmosphere is the propagation of gravity waves from the troposphere to the upper mesosphere and lower thermosphere, tunneling through an evanescent region \citep{Walt01}. Gravity waves are evanescent in neutral-stability layers, in which the energy of the waves decrease exponentially with altitude. However, if the vertical extension of the neutral region is not too large, it is possible for a significant fraction of the incident wave energy to pass through the neutral stability region. By analogy with quantum mechanics, some transfer of energy is possible via wave tunneling \citep{Suth04} through a neutral barrier. To quantify the energy transmitted from a region with a Brunt-V\"ais\"al\"a frequency $N$ through a barrier of zero Brunt-V\"ais\"al\"a frequency, the transmission $T$ is defined as

\begin{equation}
    T = \left[ 1 + \left( \frac{\sinh(k_x H )}{\sin2\theta} \right)^2 \right]^{-1} 
\end{equation}

\noindent according to \citet{Suth04}, with $k_x$ the horizontal wavenumber, $H$ the height of the barrier and $\theta$ equals to $\cos(\omega / N)$ with $\omega$ the frequency of the wave. 

For the first barrier, between 18 and 35~km, the horizontal wavelength is of the order of 200~km, $H$ is equal to 17~km, $\omega$ to 3.4~10$^{-4}$~s$^{-1}$ and $N$ to 2.3~10$^{-3}$~s$^{-1}$. For the second barrier, the convective layer between 48 and 52~km, the horizontal wavelength is of the order of 200~km, $H$ is equal to 4~km, $\omega$ to 1.3~10$^{-3}$~s$^{-1}$ and $N$ to 9.0~10$^{-3}$~s$^{-1}$. With these values, the transmission $T$ for the first layer is equal to 21~\% and to 84~\% for the second barrier. The cloud convective layer depth is closer to 10~km \citep{Tell09}, with that value the transmission drops to about 45~\%.

We conclude that, despite the presence of the two neutral-stability layers in the atmosphere of Venus, there is a significant energy transmission through these two barriers. In other words, this tunneling effect enables the orographic gravity waves to propagate towards the top Venusian clouds, where those waves are detected by Akatsuki \citep{Fuku17,Kouy17}. 

The mixed layer in the deep atmosphere between altitudes 18 and 35 km \citep{Schu80} is the thickest of the two barriers, and is the one that affects the most the vertically-propagating wave, since a fifth of the wave energy makes it through this mixed layer. Our simulations thereby provide insights into the mechanisms responsible for the propagation of the bow-shape perturbation from the surface to 35~km, left unexplained by the approach based on gravity-wave parameterization used in \citet{Nava18}. Conversely, the cloud convective layer between altitudes 48 and 52 km, thinner than the deep atmosphere mixed layer but with strong vertical plumes \citep{Lefe18}, plays a minor role and does not affect substantially the wave vertical propagation (more than 80\% of the incoming wave energy is propagating through this mixed layer). These two neutral barriers would also be an obstacle to the vertical propagation of sound waves \citep{Mart18} induced by putative Venusian seismic activity, possibly detectable via infrasound sensors on balloons \citep{Kris18} or through airglow radiation perturbation \citep{Didi18}.

\bigbreak
Saturation of a wave occurs either through critical levels (when the speed of the background horizontal flow is equal to the wave phase speed) or wave breaking through convective instability. To quantify this probability, the saturation index $S$ of \cite{Hauc87} is used 

\begin{equation}
    S = \sqrt{\frac{F_0 N}{\rho k_x |\overline{u} - c|^3}}
\end{equation}

\noindent where $F_0$ is the vertical momentum flux, $N$ is the Brunt-V\"ais\"al\"a frequency, $\rho$ the density of the atmosphere, $k_x$ the horizontal wave number $c$ the phase speed of the wave, and $\overline{\cdot}$ the average over our chosen mesoscale domain (thought to represent the large-scale component). 
When $S$ reaches values close to 1 the wave may be likely to break through critical level or saturation. For the case of Venus, the main wave is orographic and stationary with respect to the surface, with a phase speed $c$ equals to zero. The zonal wind speed is constantly increasing up to roughly 80~km of altitude, thus $|\overline{u} - c| \gg 0$ and superior to $F_0 N /k_x$, which causes~$S$ to be much smaller than~$1$, even for the largest values of stability~$N$. The saturation index only reaches values up to 1~10~$^{-1}$, and most often values of 1~10~$^{-2}$, thus the probability of saturation of the mountain waves is low. 

\begin{figure}[!ht]
 \center
 \includegraphics[width=10cm]{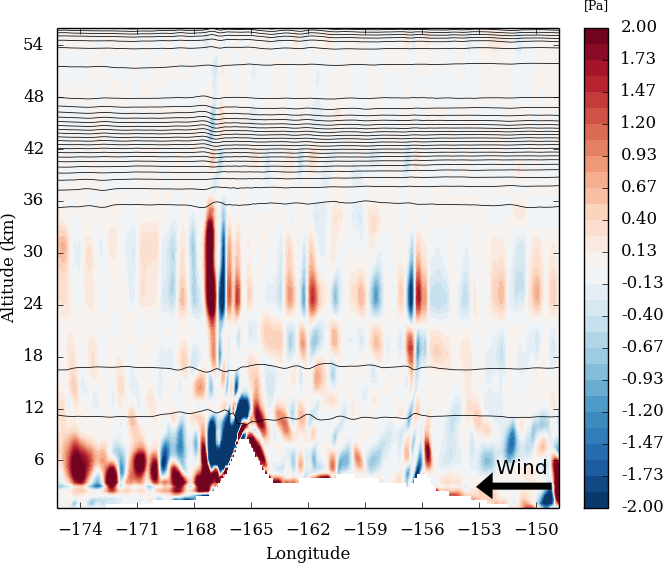}
 \caption{Instantaneous vertical cross-section of the quantity $- \rho {u' w'}$ (Pa) at 1\textdegree~of latitude, between the surface and approximately 55~km in the beginning of afternoon. Contours represent the potential temperature. Direction of the zonal is indicated with the black arrow.}
 \label{fig:34}
\end{figure}

The propagation of the waves into the two mixed layers engender a dissipation of the waves: the part of the energy that is not transferred upward through tunneling effect is deposited in those mixed layers. This deposition of momentum can be quantified by calculating the vertical component of the Eliassen-Palm momentum flux \citep{Andr87} that is equal to $- \rho \overline{u' w'}$ where $\rho$ is the density and $u'$ and $w'$ the perturbations of the zonal wind $u$ and vertical wind $w$ (with respect to the mean defined as~$\overline{\cdot}$). A postive value of $u'$ represents an eastward velocity perturbation and a positive value of $w'$ represents an upward velocity perturbation and therefore a positive value of $- \rho \overline{u' w'}$ represents an upward transport of westward momentum. The quantity displayed at Figure~\ref{fig:34} is an instantaneous snapshot of the quantity $- \rho {u' w'}$. This flux is around 10 times larger than the strongest measured on Earth in Antarctica \citep{Jewt15}. This is partly due to the fact that the density of the Venusian atmosphere is 65 times larger than the Earth atmosphere. Momentum is transported by the waves near the surface (where density is the highest), but also in the mixed layer around 30~km of altitude (where wave perturbations are strong). The trapped lee waves also transport momentum horizontally, while the vertical transport of momentum due to leakage is negligible. The momentum flux~$-\overline{\rho u' w'}$ is the drag coefficient \citep{Smit79} calculated in sub-grid-scale parameterizations, such as the one used in the GCM runs by \citet{Nava18}, who employed the orographic-wave parameterization of \citet{Lott97}. In this parameterization, \citet{Nava18} had to adopt values of 2~Pa for the threshold of the gravity wave mountain stress, and 35~km for the initial altitude of deposition of this mountain stress, to reproduce the bow-shape waves observed by Akatsuki. Our mesoscale modeling of the Venusian mountain waves, in which the propagation of gravity waves is resolved from the surface to 100~km, therefore validates the assumption of the \citet{Nava18} study, both for the amplitude of the stress and the altitude of the stress deposition (a maximum of momentum is deposited by the waves at this altitude), and shows that they are physically based. 

\begin{figure}[!ht]
 \center
 \includegraphics[width=14cm]{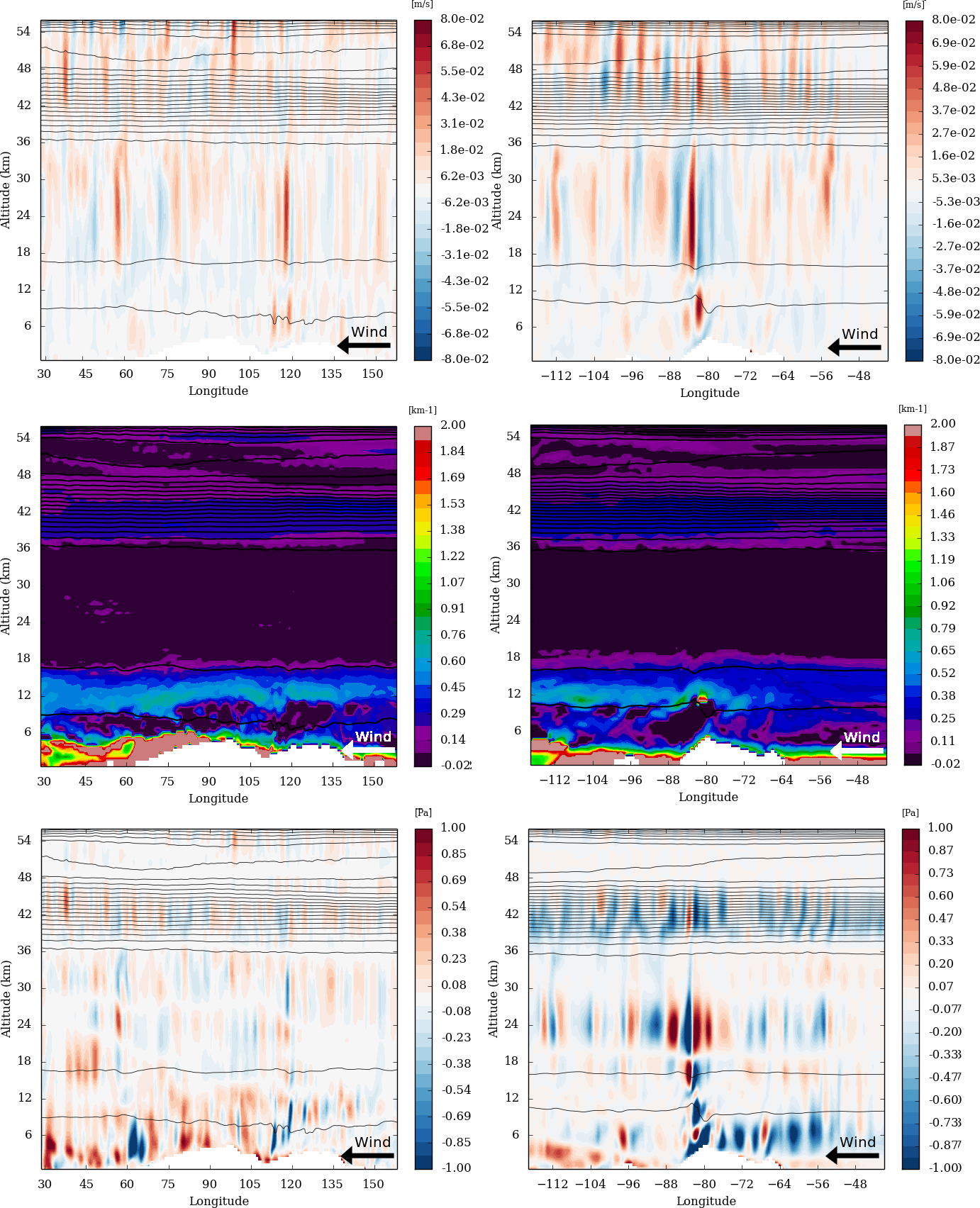}
 \caption{Vertical cross-section of the vertical wind (m~s$^{-1}$) on top panel and vertical cross-section of the Scorer parameter (km$^{-1}$) on middle panel and vertical cross-section of vertical momentum flux (Pa) on bottom panel for Aphrodite Terra on left column and Beta Regio on right column. Contour is potential temperature. Direction of the zonal is indicated with arrows.}
 \label{fig:35}
\end{figure}

What we describe here for Alta Regio extends to the other regions considered in this study. Figure~\ref{fig:35} shows the vertical cross-section of the vertical wind, Scorer parameter and momentum flux for Aphrodite Terra (left) and Beta Regio (right). As for Atla Regio, the major terrain elevations generate gravity waves and the flow is in a non-linear regime. The two layers of low static stability are also present in those regions, which entails the generation of trapped lee waves. The amplitude of the vertical wind is smaller than in the Atla case due to lower slopes. Thus, the amplitude of the momentum flux is of the same order of magnitude, albeit slightly smaller. The vertical extent of the two mixed layers is very similar to Atla Region, in both the Aphrodite Terra and Beta Regio cases, but the horizontal wavelengths are slightly larger and resulting in a small decrease of the tunneling transmission $T$ but in the same order of magnitude, about 20~\% for the first barrier (18-35 km altitudes) and 80~\% for the second barrier (48-52 km altitudes).

\section{Aphrodite Terra}
\label{Sec:AT}

\begin{figure}[!ht]
 \center
 \includegraphics[width=10cm]{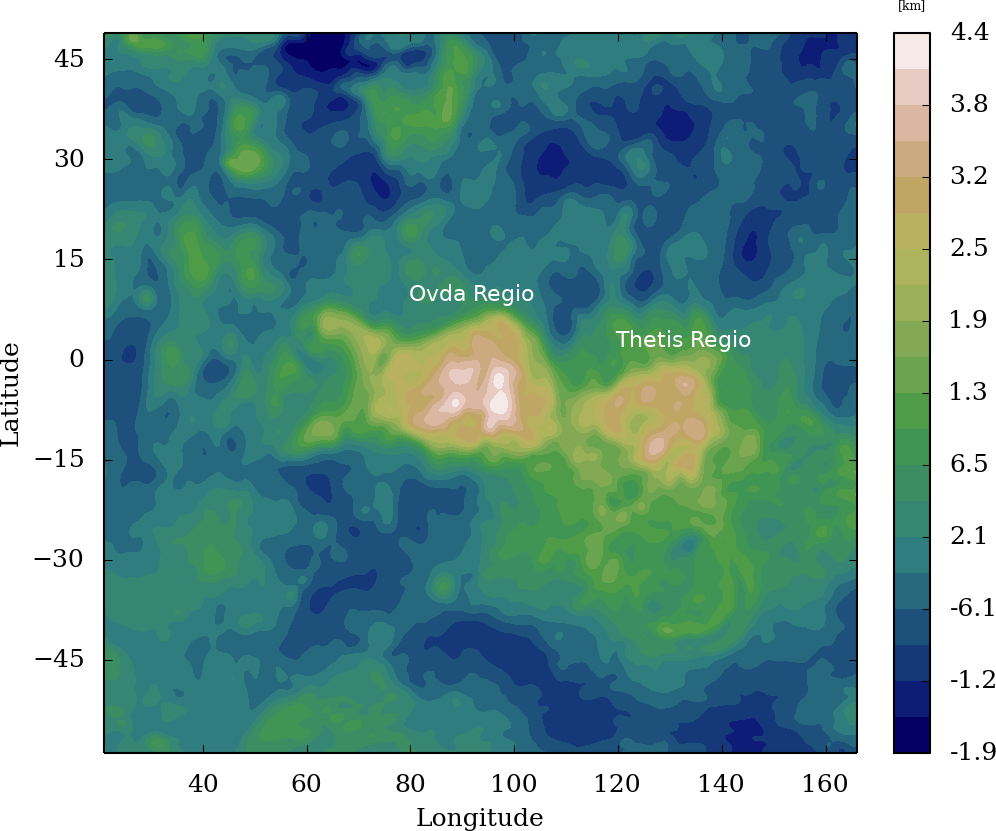}
 \caption{Elevation map (km) of the selected domain for Aphrodite Terra with a resolution of 40~km.}
 \label{fig:411}
\end{figure}

Figure~\ref{fig:411} shows the selected domain for Aphrodite Terra with a resolution of 40~km, Ovda Terra is visible between 60 and 100\textdegree~of longitude where the main bow-shape gravity waves have been observed by \textit{Akatsuki}. 

\textit{Akatsuki} LIR measurements consist of a mean of the brightness temperature between 8 and 12 $\mu$m to which is applied a high-pass filter \citep{Fuku17,Kouy17,Tagu07}. To be able to compare the outputs of the model with the observations, we have to analyze in the model the deformation of the cloud top by the gravity waves. We consider that the characteristic timescale of the wave propagation is smaller than the radiative timescale, so that the deformation is adiabatic; it follows that potential temperature can be used as a tracer for the deformation of the cloud top by waves. By choosing one value for potential temperature to set a relevant material surface, a corresponding temperature map can be reconstructed, as well as a map of the corresponding altitude. To compare the waves resolved by the model to the ones observed by the \textit{Akatsuki} spacecraft, anomaly temperature is calculated for potential temperature surface between 55 and 80~km, corresponding to a range of potential temperature from 800 to 1300~K, with a gaussian weighting function mimicking the LIR's weight function. Then the temperature perturbations are vertically averaged. A high-pass gaussian filter is applied to filter out the global dynamics component \citep[similarly to what is done in the maps derived from Akatsuki observations][]{Fuku17,Kouy17}, and a low-pass gaussian filter is also applied to remove the small-scale features smaller than a few tens of kilometer (i.e. approaching the effective horizontal resolution of our mesoscale simulations for Venus). Figure~\ref{fig:422} presents the associated residuals of the temperature anomaly after filtering. In the end, temperature maps such as Figure~\ref{fig:421} obtained from our mesoscale modeling by this method can be directly compared to the \textit{Akatsuki} observations in \citet{Fuku17} and \citet{Kouy17}.  

\subsection{Resulting bow-shape wave}

\begin{figure}[!ht]
 \center
 \includegraphics[width=16cm]{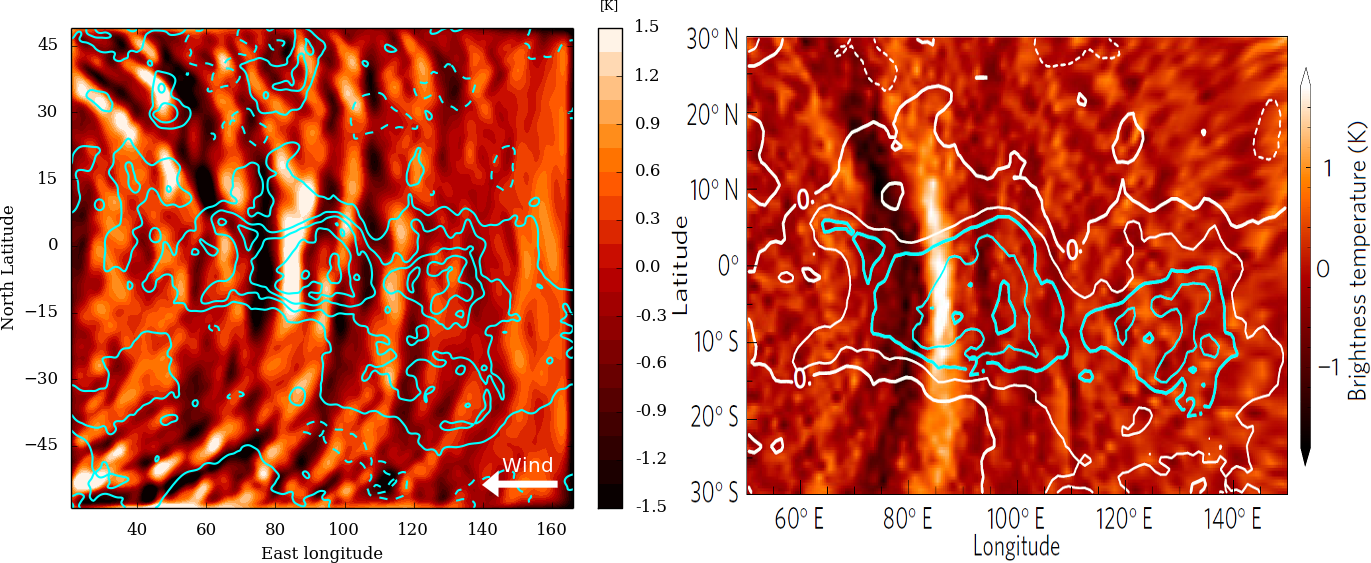}
 \caption{Left : Temperature anomaly (K) at the top of the cloud in late afternoon. Cyan contours show the topography every 1~km (Fig.~\ref{fig:411}). Direction of the zonal is indicated with the white arrow. Right : Temperature anomaly (K) at the top of the cloud observed by \textit{Akatsuki}/LIR adapted from \cite{Fuku17}. White and cyan lines are topography.}
 \label{fig:421}
\end{figure}

\begin{figure}[!ht]
 \center
 \includegraphics[width=16cm]{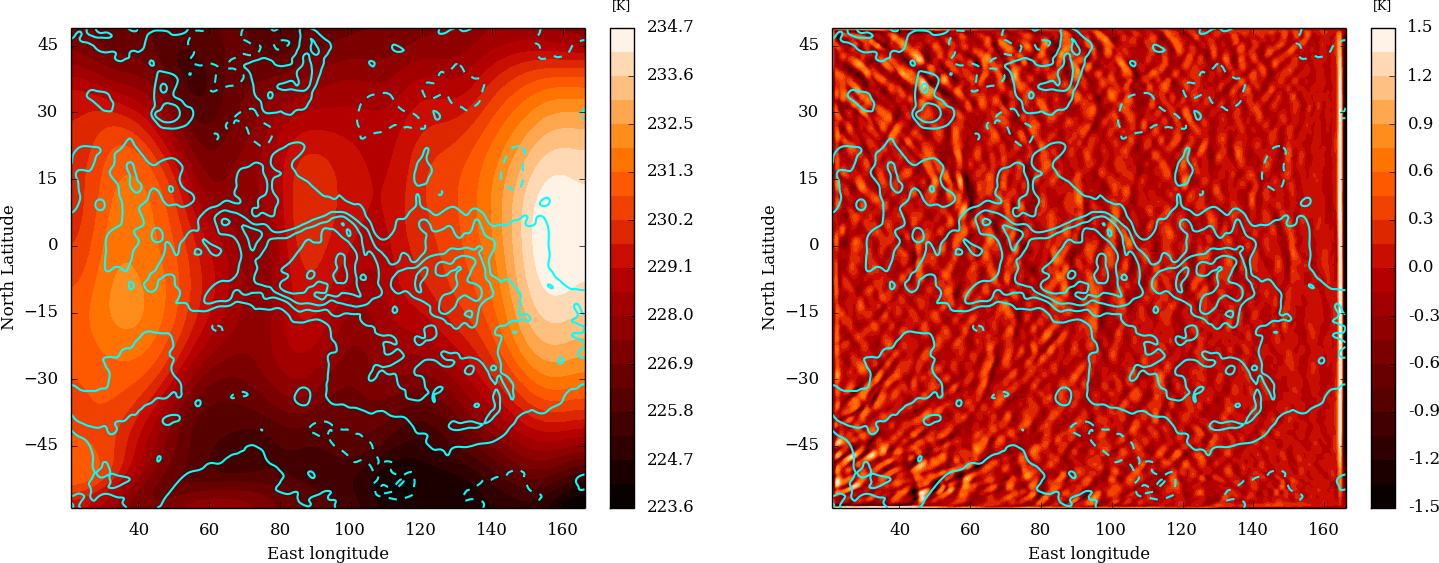}
 \caption{The scene displayed in this figure is similar to Figure~\ref{fig:421}. The low-pass filter signal (left panel) representing the global dynamics component has a maximum around 160\textdegree~of longitude, where the local time is close to noon. The high-pass filter signal (right panel) shows small-scale wave features not visible on Akatsuki LIR images \citep{Kouy17} but discernible on UV images \citep{Kita19}. Cyan contours show the topography (Fig.~\ref{fig:411}).}
 \label{fig:422}
\end{figure}

Figure~\ref{fig:421} shows the temperature anomaly modeled by our LMD Venus mesoscale model at the top of the cloud above Aphrodite Terra, close to the evening terminator. Cyan contours show the topography. The topography-induced perturbation at the top of the cloud expands from -40 to 40\textdegree~of latitude with a bow-shaped morphology, resembling the measured signal by Akatsuki. Above Ovda Terra, the positive and negative anomaly measured by LIR \citep{Fuku17} is reproduced with a similar amplitude, \textpm 2~K. Outside of the bow-shape wave small-scale waves features are visible, similar structures are visible in the UV images \citep{Kita19}. This orographic gravity wave induces a deformation of about 600~m of the cloud top altitude. Temperature anomalies induced by mountain waves are visible in the middle and upper cloud layers whereas no temperature anomaly is discernible in the lower cloud layer due to the presence of the convective layer. This could explain the fact that VEx/VIRTIS measured on the nightside stationary waves in the upper cloud layer but not in the lower cloud layer \citep{Pera17}.

The divergence of the momentum flux indicates the acceleration (or equivalently the force per unit mass) caused by gravity waves on the mean flow when they break or encounter a critical level \citep{Frit03}. This acceleration can be calculated by the equation
\begin{equation}
    \frac{\partial \overline{u}}{\partial t} = \frac{1}{\rho}\frac{\partial}{\partial z} \rho \overline{u' w'}
\end{equation}
\noindent Observations indicate that around cloud top, for altitudes above 67~km, there is a deceleration of the zonal wind. 

\cite{Bert16} calculated that an deceleration of 13~m~s$^{-1}$ per Venus day was necessary to explain longitudinal shift of zonal wind patterns. The \textit{Akatsuki} spacecraft measured a longitudinal variability of the zonal wind around cloud top between 4 and 12~m~s$^{-1}$ at the Equator \citep{Hori18} with no correlation with topography. The deceleration induced by the bow-shape waves resolved by our LMD Venus mesoscale model, integrated over a Venus day, reaches values around 3~m~s$^{-1}$ consistent with the lower range of estimates based on \textit{Akatsuki} measurements. This value is smaller than the computations in \citet{Bert16} (and the higher range of Akatsuki estimates), which tends to indicate that the deceleration of zonal wind at cloud top cannot be explained solely by the impact of bow-shaped gravity waves.

\subsection{Variability with local time}

\begin{figure}[!ht]
 \center
 \includegraphics[width=10cm]{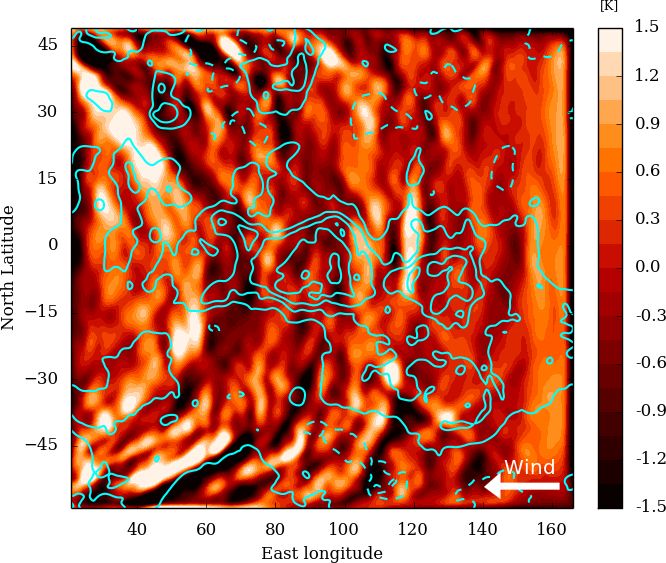}
 \caption{Temperature anomaly (K) at the top of the cloud 3.5 Earth day after Figure~\ref{fig:421}. Cyan contours show the topography every 1~km (Fig.~\ref{fig:411}). Direction of the zonal wind is indicated with the white arrow.}
 \label{fig:431}
\end{figure}

The temperature anomaly of 2~K at the top of the cloud shown in Figure\ref{fig:421} is the maximum signal obtained. This wave is visible in the model with an amplitude superior to 0.5~K for about 10 Earth days, against approximately 14 Earth days for the \textit{Akatsuki} observation (between 14h et 17h in local time). Figure~\ref{fig:431} shows Temperature anomaly (K) at the top of the cloud 3.5 Earth day after Figure~\ref{fig:421}, the bow-shape wave above Ovda Regio is still visible but with smaller amplitude meanwhile a bow-shape wave above Thetis Regio is discernible similarly to LIR observations \citep{Kouy17}.

Additional simulations were performed at midnight and noon to study the variability of the waves at cloud top. At midnight, no significant wave (amplitude superior to 0.5~K) is observed. At noon, transient waves are visible with amplitude smaller than 1~K. The variation of the surface wind along the day, about 1~m~s$^{-1}$ above Ovda Terra, is too small to explain this variability. This indicates that the near-surface conditions responsible for the emission of orographic gravity wave are not changing significantly along the day. The diurnal variability of conditions responsible for the propagation of orographic gravity waves towards high altitude is more likely to explain the observed variability with local time of the bow-shaped waves.

\begin{figure}[!ht]
 \center
 \includegraphics[width=16cm]{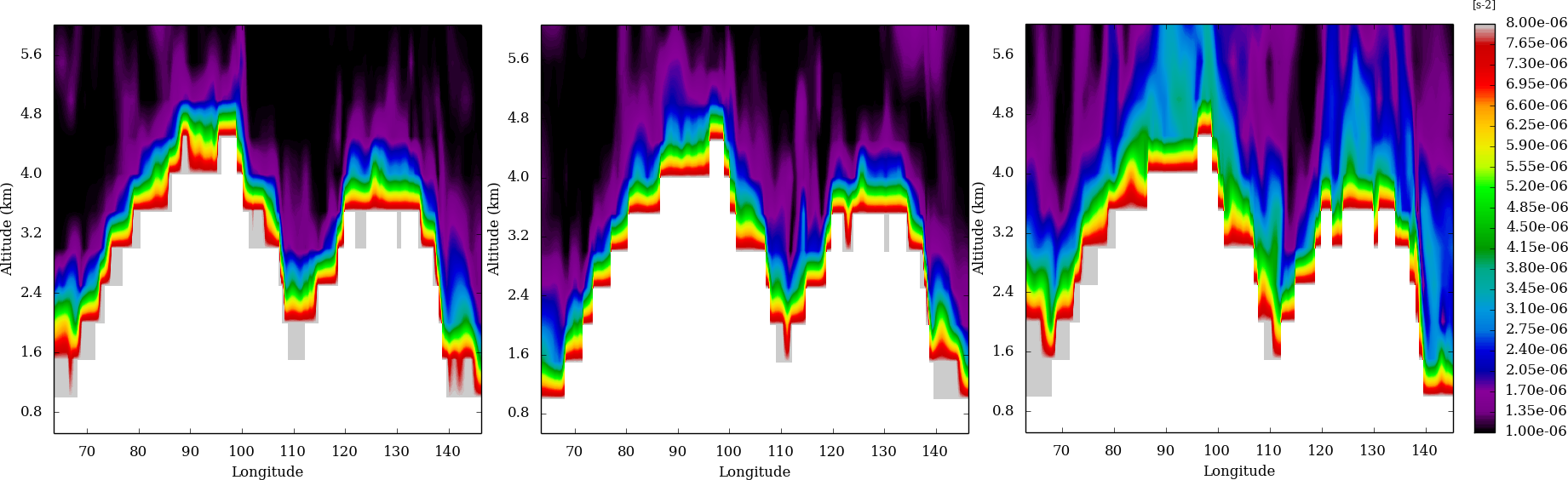}
 \caption{Square of the Brunt-V\"ais\"al\"a frequency (s$^{-2}$) of the atmosphere from the surface to 6~km for midnight (left), noon (middle) and late afternoon (right) for Aphrodite Terra.}
 \label{fig:432}
\end{figure}

Figure~\ref{fig:432} shows the square of the Brunt-V\"ais\"al\"a frequency (s$^{-2}$) above Aphrodite Terra between the surface and 6~km of altitude for the three local times : midnight (left), noon (center), late afternoon (right). Close to the surface the atmosphere is very stable, but this stability is decreasing with altitude by an order of magnitude. This decrease is different depending on the local time. At midnight, the static stability shows a strong gradient with altitude. At noon this gradient is smaller, and even smaller in late afternoon with constant value of 3~10$^{-6}$~s$^{-2}$ over more than one kilometer. This slow decrease has consequences over the Scorer parameter, it is superior to 2~km in the first 8~km in late afternoon, against only in the first 3~km at other local times. This strong value of the Scorer parameter over several kilometers favors the vertical propagation of the gravity waves. This is correlated to the vertical momentum flux: at night the amplitude is about 10$^{-2}$~Pa, while it is about 10$^{-1}$~Pa at noon. Meanwhile, the very-near surface (first hundred meters) stability is consistent with \cite{Nava18} study : the atmosphere is more stable at night, but with a smaller local-time variability. This behavior plays a role in the generation of the waves, while the first 4 to 5 km affect the vertical propagation.

\section{Atla and Beta Regios}

\subsection{Atla Regio}

\begin{figure}[!ht]
 \center
 \includegraphics[width=16cm]{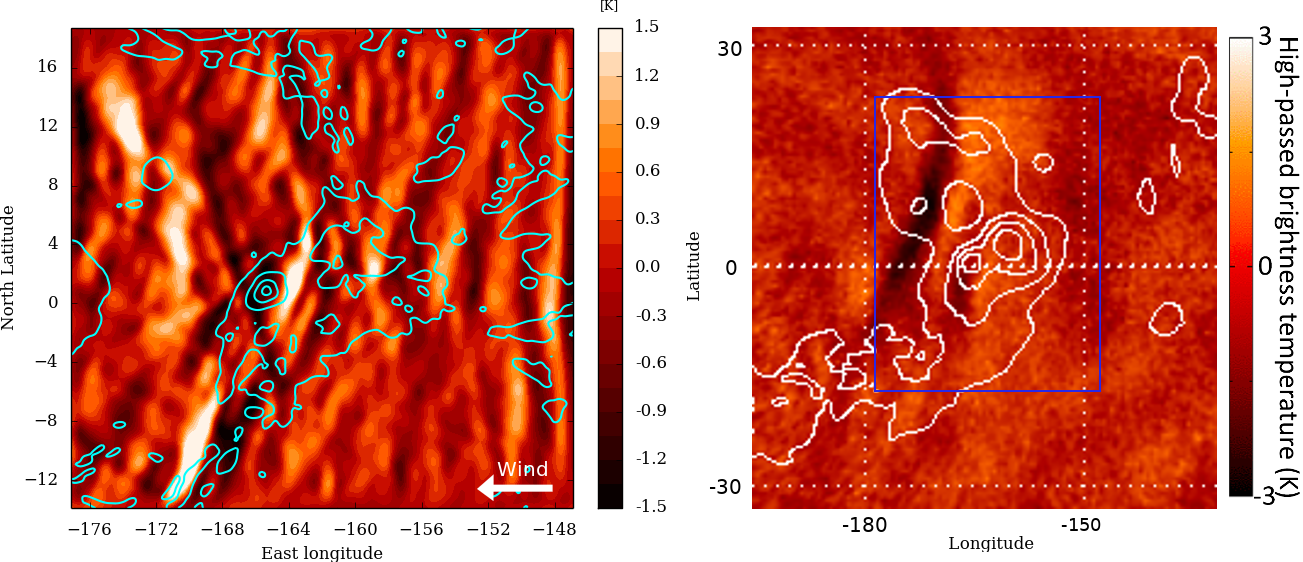}
 \caption{Left : Temperature anomaly (K) at the top of the cloud in the afternoon. Cyan contours show topography every 1.6~km (Fig.~\ref{fig:31}). Direction of the zonal is indicated with the white arrow. Right : Temperature anomaly (K) at the top of the cloud observed by \textit{Akatsuki}/LIR adapted from \cite{Kouy17}. White and cyan lines are the topography. Blue lines on the right panel show the mesoscale domain.}
 \label{fig:511}
\end{figure}

The left panel of Figure~\ref{fig:511} shows the resulting temperature anomaly at cloud top for the Atla Terra (Fig.~\ref{fig:31}) simulation, using the same methodology as for Aphrodite Terra. Two waves are visible, one above the main mountain Maat Mons at -168\textdegree~of longitude and an other over Ozza Mons at -163\textdegree~of longitude. The amplitude of these waves is of about \textpm1.5~K for the first one and about \textpm1~K for the last one. The latitudinal expansion is about 20\textdegree. 

LIR observed one main wave above Ozza Mons with an amplitude of \textpm2~K and an extension of 30\textdegree~as well as a second wave of a few degrees of latitude associated with Maat Mons \citep{Kouy17}. The main resolved wave is similar to the observations, with however sharp morphology that may be due to the abrupt terrain elevation; the second wave is more extended in latitude than in the observations. These waves are obtained for the beginning of the afternoon when the amplitude is maximum, whereas LIR observed the maximum in the middle of the afternoon. The waves resolved by the model are symmetrical to the obstacle, meaning the latitudinal expansion towards the north and towards the south is the same, while LIR observed non-symmetrical waves attributed to the complex terrain. 

The generation and propagation of the waves are sensitive to the wind and stability conditions close to the surface and therefore to physical representations within the model, for example the radiative transfer. The Venus GCM may not represent fully accurate surface conditions, especially for winds, as well as their evolution with time. These differences could explain that, while the overall bow-shaped wave is reproduced in our Venus mesoscale simulation, some discrepancies with the observations do exist.

\subsection{Beta Regio}

\begin{figure}[!ht]
 \center
 \includegraphics[width=10cm]{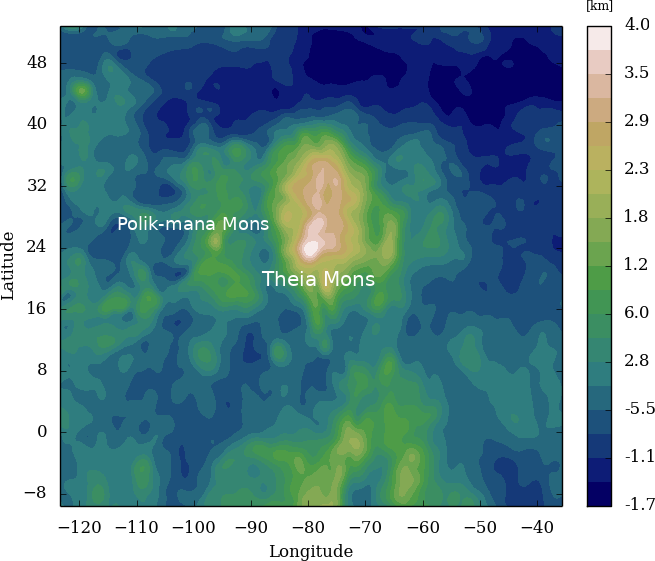}
 \caption{Elevation map (km) of the selected domain for Beta Regio with a resolution of 30~km.}
 \label{fig:521}
\end{figure}

\begin{figure}[!ht]
 \center
 \includegraphics[width=16cm]{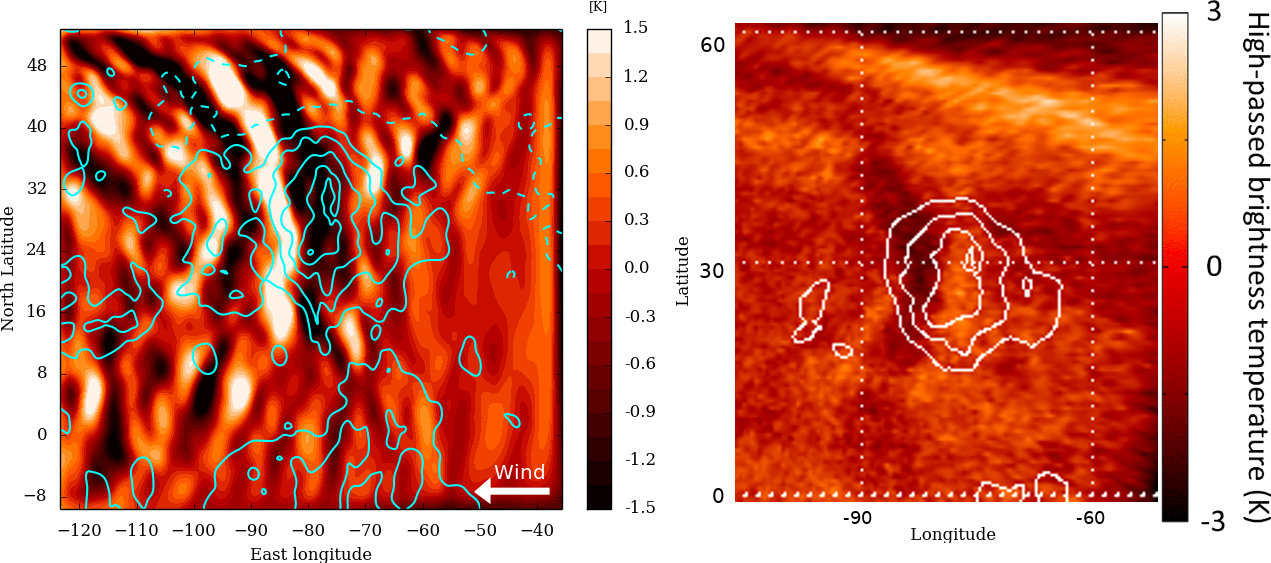}
 \caption{Left : Temperature anomaly (K) at the top of the cloud in late afternoon. Cyan contours show topography every 850~m (Fig.~\ref{fig:521}). Direction of the zonal is indicated with the white arrow. Right : Temperature anomaly (K) at the top of the cloud observed by \textit{Akatsuki}/LIR adapted from \cite{Kouy17}. White and cyan lines are the topography.}
 \label{fig:522}
\end{figure}

Figure~\ref{fig:521} shows the selected domain Beta Regio with a resolution of 30~km, where the highest elevation point is Theia Mons at -80\textdegree~of longitude and 30\textdegree~of latitude. The temperature anomaly at the top of the cloud is shown in left panel of Figure~\ref{fig:522}. Two distinct stationary waves are visible, a main one above Theia Mons and a smaller one above Polik-mana Mons at 30\textdegree~of latitude and -100~\textdegree. These two waves are observed by \textit{Akatsuki}/LIR with the same relative size visible in the right panel of Figure~\ref{fig:522}. The bow shape is visible with an amplitude of \textpm1.5~K. A difference with previous anomaly is the asymmetry of the waves, which is more developed towards the North. This non-symmetrical shape is not clearly observed above Beta Regio in LIR images \citep{Kouy17} but visible in UV images \citep{Kita19}. The latitude of the region, 30\textdegree, may have an impact on the morphology of the wave.

\section{Polar regions}

\begin{figure}[!ht]
 \center
 \includegraphics[width=16cm]{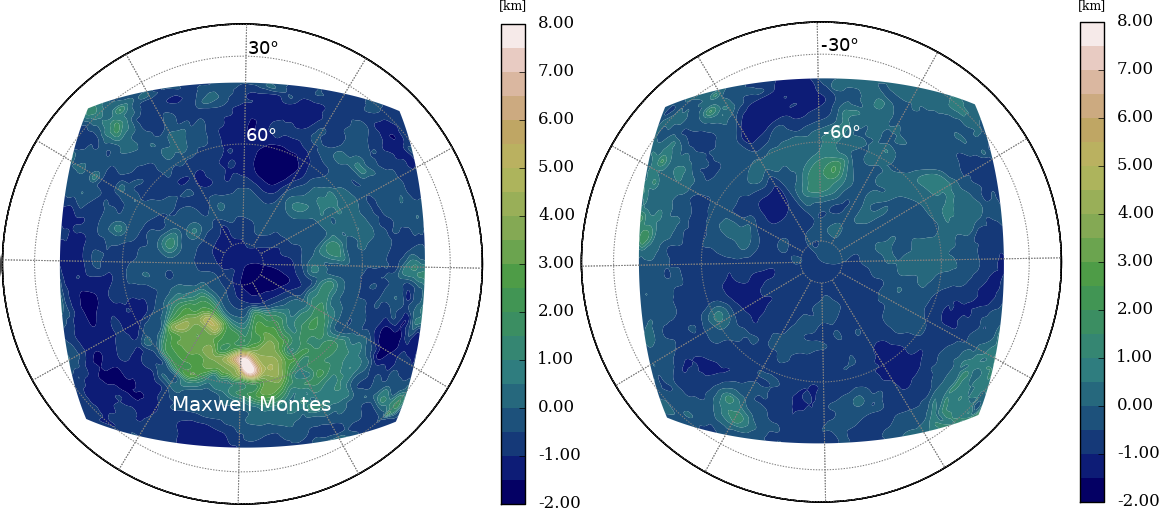}
 \caption{Elevation map (km) of the selected domains for the North pole (left) and the South pole (right) regions with a resolution of 40~km.}
 \label{fig:61}
\end{figure}

The highest topographical point is Maxwell Montes peaking at approximately 10~km at 65\textdegree N latitude. This is a site of interest in orographic gravity waves on Venus. Unfortunately, the orbit of \textit{Akatsuki} is equatorial and therefore the spacecraft cannot observe the atmospheric activity in polar regions; \textit{Venus Express} had an elliptical polar orbit suited to study the south pole, but not the north pole. To study the generation of orographic waves above Maxwell Montes, and more generally in the polar regions, mesoscale simulations can be performed with polar stereographic projection over both poles for comparison. The chosen domains for North (right) and South pole (left) are shown in Figure~\ref{fig:61}, with a resolution of 40~km for both domains. Recent improvements of the IPSL Venus GCM used to provide initial and boundary for our Venus mesoscale model, especially in the polar regions, ensure a correct large-scale forcing with the inclusion of the cold collar and mid-to-high-latitude jets \citep{Gara18}.

\begin{figure}[!ht]
 \center
 \includegraphics[width=16cm]{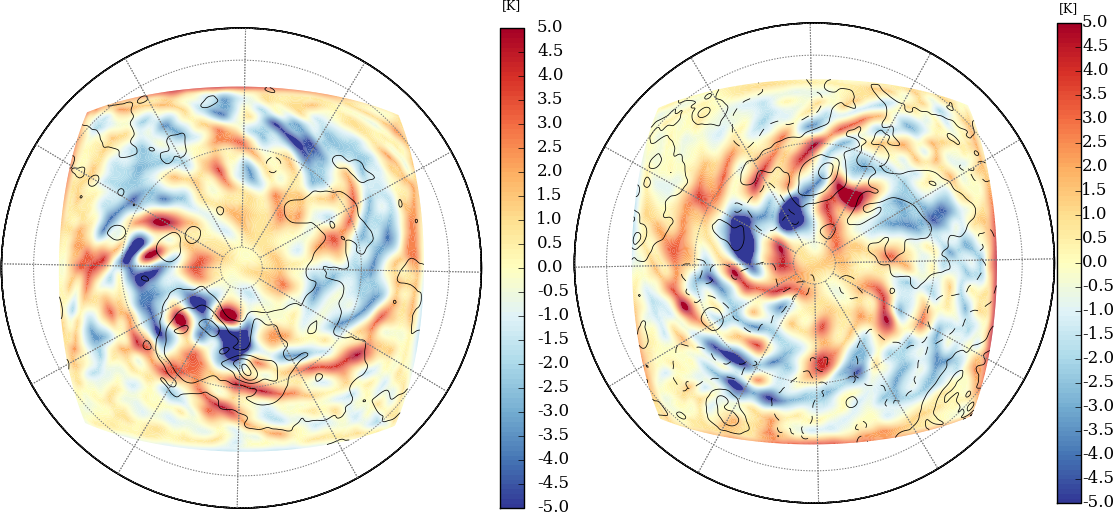}
 \caption{Temperature anomaly (K) at the top of the cloud for the North pole (left) and the South pole (right) regions. Black contours show topography every 1.6~km for the North pole and every 800~m for the South pole (Fig.~\ref{fig:61}).}
 \label{fig:62}
\end{figure}

Figure~\ref{fig:62} shows the temperature anomaly at the top of the cloud. Transient waves rotating around the planet are visible in both polar regions down to $\pm$ 60\textdegree~with an amplitude larger than 5~K. The wave structure and amplitude are similar for both regions meaning that the underlying topography has hardly any impact on the temperature anomaly at cloud top. These waves resemble the planetary streak structure visible in the \textit{Akatsuki}/IR2 images \citep{Lima18} and reproduced with GCM modelling \citep{Kash19}.

\begin{figure}[!ht]
 \center
 \includegraphics[width=10cm]{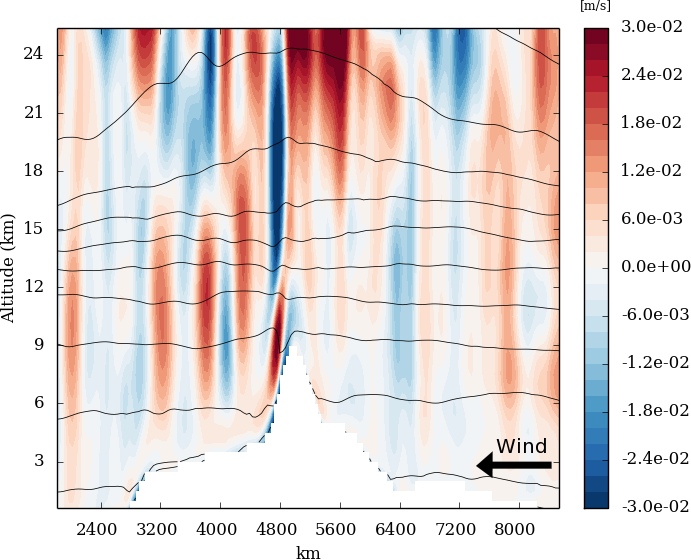}
 \caption{Vertical cross-section of the vertical wind (m~s$^{-1}$) between the surface and approximately 55~km. Please note that the cross-section is not at a constant latitude. Contours represent the potential temperature.}
 \label{fig:63}
\end{figure}

However, mountain waves are generated by Ishtar Terra. The vertical velocity above Maxwell Montes is shown in Figure~\ref{fig:63}. These waves propagate vertically and encounter the two low static stability regions but do not yield any stationary temperature anomaly at cloud-top. The two barriers are thicker than that for the other cases, 22 against 17~km for the first one and 7 against 4~km for the convective layer. The increase of the cloud convective layer thickness is expected from observations \citep{Tell09,Haus14} and modelling \citep{Imam14,Lefe18}, yet few observations have been performed regarding the mixed layer in the deep atmosphere, especially regarding the variation of depth with latitude and local time. Only the temperature profiles of the Pioneer Venus probes \citep{Seiff80} would be available for such an analysis. 

As is discussed in section~\ref{Sec:gwg}, the two mixed layers in the lower atmosphere of Venus act as barriers for gravity waves. For the first barrier (altitudes 15-35~km) with $\omega$ equals to 2.3~10$^{-4}$~s$^{-1}$, $N$ to 1.6~10$^{-3}$~s$^{-1}$ and a horizontal wavelength of 200~km, the transmission is only of 13~\%. For the second barrier (altitudes 45-55~km) with $\omega$ equals to 1.1~10$^{-3}$~s$^{-1}$, $N$ to 7.5~10$^{-3}$~s$^{-1}$ and the same horizontal wavelength, the transmission drops to 63~\%. The orographic wave is therefore strongly affected by propagating through the two mixed layers (especially the lowermost one). As a result, the cloud top temperature anomaly induced by gravity waves in polar regions is smaller than the ones in the equatorial regions, and negligible against the cloud top transient waves. The large amplitude of the transient polar waves could prevent the visibility of the comparatively-low stationary gravity-wave perturbation and explain the non-observation of stationary waves in \textit{Venus Express}/VIRTIS data at latitudes above 65\textdegree S \citep{Pera17}. The saturation of the waves is here again unlikely, so contrary to mixed layers acting as efficient barriers, it does not seem a plausible mechanism to explain the lack of observed gravity-wave stationary wave in polar regions.

\section{Conclusion}

We present here the first mesocale model applied to Venus, in which the WRF dynamical core is coupled the set of physical parameterizations developed for the IPSL Venus GCM, notably the radiative transfer to compute the solar and IR rates during the simulations. Focusing on three areas of interest, the model resolves bow shape stationary waves with an amplitude around 1.5~K. The lifetime of the waves is about ten Earth days. The maximum amplitude of the waves are in the afternoon, close to the terminator for Aphrodite Terra, earlier for the two another cases. The characteristics and morphology of the waves observed by Akatsuki are overall well reproduced by the model. 

Theses gravity waves are generated by the large-scale flow forced to go over Venus' mountains. Propagating vertically, those orographic gravity waves encounter two layers of low static stability, the mixed layer (20-32~km) and the convective layer (47-55~km) where some energy is transmitted through the layers via tunneling phenomenon. The deep mixed layer is the most critical barrier for the wave, and comprehensive studies about the variability of this layer with latitude and local time would improve the understanding of the propagation of the mountain waves. The presence of these two layers generate trapped lee waves that propagate horizontally with some vertical leakage due to their large horizontal wavelength. 

The variations of the wave characteristics with local time is imputed to the stability of the atmosphere close to the surface. In the afternoon, the first 4-5~km of the atmosphere are more stable that at night or at noon, which favors the vertical propagation of the waves. 

The waves extract momentum from the surface with a flux around 2~Pa and deposit this momentum at around 30~km. Values of parameters used in the orographic parameterization of \citet{Nava18} are therefore be confirmed by our mesoscale model which resolves the emission and propagation of gravity waves from the surface to 100-km altitude. At the cloud top, the waves induce a deceleration of several meters per second over a Venus day, consistent with \textit{Akatsuki} measurements.

Temperature anomalies above Venus' polar regions have also been explored with our mesoscale model. Transient waves rotating around the poles dominate the signal and the influence of the underlying topography is not noticeable. However, mountain waves are generated above Ishtar Terra -- similarly to the waves generated by the other main mountains at lower latitudes. The increase of the deep atmosphere neutral barrier thickness strongly affects the amplitude of the orographic waves, which become negligible against the circumpolar transient waves. 

Surface wind and the two mixed layers are key factors for the generation and propagation of mountain waves, a sensitivity study should be therefore considered in future work. Such study will imply complex changes in the IPSL Venus GCM dynamics and cloud model.

Additional simulations may be performed in the future for several other regions of interest, like Pheobe Regio at the Equator with a complex bow-shape morphology witnessed by Akatsuki, Gula Mons and Bell Regio, two sharp mountains at respectively 20\textdegree N and 30\textdegree N and Imdr Regio composed of two sharp mountains at 45\textdegree S to investigate the influence of the impact of the morphology of the mountain, of the conditions near the surface and of the latitude on the temperature anomaly. Moreover the observations of stationary waves by VEx/VIRTIS in the southern hemisphere in the nightside with no apparent dependence with the local time \citep{Pera17} are features to be investigated.

The LMD Mesoscale model for Venus based on WRF \citep{Skam08} can be applied to the study of several other regional-scale phenomena, in particular slope winds, which were studied on Earth \citep{Brom01} and on Mars \citep{Spig11}, and play a key role in the vertical extension of the planetary boundary layer convection \citep{Lebo18} and local meteorology in general.

Implementation of the photochemistry and microphysics schemes developed at IPSL in the Venus mesoscale model is planned to investigate the influence of mountains on the chemistry and the cloud formation. Explosive volcanism could explain SO$_2$ anomaly at the top of the cloud \citep{Espo88}: the study of the vertical transport and the impact of the waves on SO$_2$ \citep{Glaz99} and other volatile like water \citep{Aire15} could constrain this phenomenon, especially above Atla Regio where hot spots have been observed with \textit{Venus Express} suggesting active volcanism \citep{Shal15}. Atmospheric variability on its own, without the need to invoke active volcanism, could also explain the observed SO$_2$ variability \citep{Marc13}.

\section*{Acknowledgements}
This work was granted access to the High-Performance Computing (HPC) resources of Centre Informatique National de l’Enseignement Supérieur (CINES)
under the allocations n°A0020101167 and A0040110391 made by Grand Équipement National de Calcul Intensif (GENCI). Simulation results used to obtain the figures in this paper are available in the open online repository https://figshare.com/s/0751293c8bbb6925a655. Full simulation results performed in this paper are available upon reasonable request (contact: maxence.lefevre@lmd.jussieu.fr). The authors acknowledge Riwal Plougonven 
and Scot Rafkin for insightful discussions on gravity waves and mesoscale modeling. The authors thank the two anonymous reviewers that helped improve the quality of the paper. The authors thank Toru Kouyama for Akatusuki/LIR images. ML and SL acknowledge financial support from Programme National de Plan{\'e}tologie (PNP).

\end{document}